\documentclass[10pt, conference, compsocconf]{IEEEtran}
\usepackage{footnote}
\usepackage{subfigure}
\usepackage{float}
\usepackage{multirow}
\usepackage{graphicx}
\usepackage{url}

\usepackage{breakurl}
\usepackage[breaklinks]{hyperref}

\usepackage[cmex10]{amsmath}
\usepackage{amsmath}
\usepackage{amssymb}
\usepackage{algorithm}
\usepackage{algorithmic}
\usepackage{xcolor}

\def\IEEEbibitemsep{0pt plus .5pt}

\begin{document}

\bstctlcite{IEEEexample:BSTcontrol}
%
\title{Service-Oriented Architecture for Drone-based Multi-Package Delivery}



\author{
\IEEEauthorblockN{Babar Shahzaad}
\IEEEauthorblockA{\textit{School of Computer Science} \\
\textit{The University of Sydney}\\
 Australia \\
babar.shahzaad@sydney.edu.au}

\and
\IEEEauthorblockN{Athman Bouguettaya}
\IEEEauthorblockA{\textit{School of Computer Science} \\
\textit{The University of Sydney}\\
Australia \\
athman.bouguettaya@sydney.edu.au}

}


%


\maketitle

\begin{abstract}
We propose a novel service-oriented architecture for drone-based multi-package delivery. The proposed architecture provides a high-level design for deploying a skyway network in a city for the effective provisioning of drone-based service delivery. A graph-based heuristic is proposed to reduce the search space for optimal service selection in the skyway network. We then find an optimal solution using the selected drone services under a range of constraints. Experimental results demonstrate the efficiency and effectiveness of our proposed graph-based heuristic approach in terms of execution time and delivery time.

\end{abstract}

%
\IEEEpeerreviewmaketitle

\section{Introduction}

Drones are a new type of IoT device that fly in the sky without a crew on board \cite{9666755}. Drones offer multi-fold advantages including fast, convenient, and cost-efficient services in various sectors \cite{10.1007/978-3-030-33702-5_28}. The key sectors include agriculture, surveillance, healthcare, shipping, and shopping \cite{8795473}.
During the COVID-19 pandemic, drones have been widely used to monitor social distancing, perform aerial spraying, and deliver goods \cite{https://doi.org/10.1111/1758-5899.13007}. Several countries have used drones for safe and contactless deliveries during lockdowns \cite{KIM2021102758}.
Big players such as Google and Amazon are testing the drones for package delivery \cite{10.1007/978-3-030-91431-8_30}. Amazon claims that 86\% of the company's items are less than five pounds (2.27 kg) and can be delivered by drones \cite{SACRAMENTO2019289}. There are three key \emph{beneficiaries} of delivery services using drones (1) consumers, (2) suppliers of goods, and (3) transport companies.

The \textit{service paradigm} is a powerful and elegant mechanism to unlock the full potential of a drone and abstract its capabilities \cite{DEVRIEZE2011637,9707164}. We model the delivery of packages using a drone as a \textit{drone service}. In this respect, the functional attribute of the service describes the {\em transport of a package} in a predefined {\em skyway} network from one node to another. The non-functional (i.e., \textit{Quality of Service} (QoS)) attributes include the flight range and payload and battery capacities of a drone. A {\em skyway network} is constructed by connecting the nodes in a geographical area where each node represents take-off (e.g., warehouse) and landing (e.g., customer's building) stations \cite{9767410,8818436}. The nodes in this network may be \textit{delivery target} to drop off the packages and/or \textit{recharging station} to land and recharge the drone. The package delivery along a skyway segment represents an \textit{individual drone service} where each service operates under a range of constraints.

\begin{figure}
    \centering
    \includegraphics[width=0.4\textwidth]{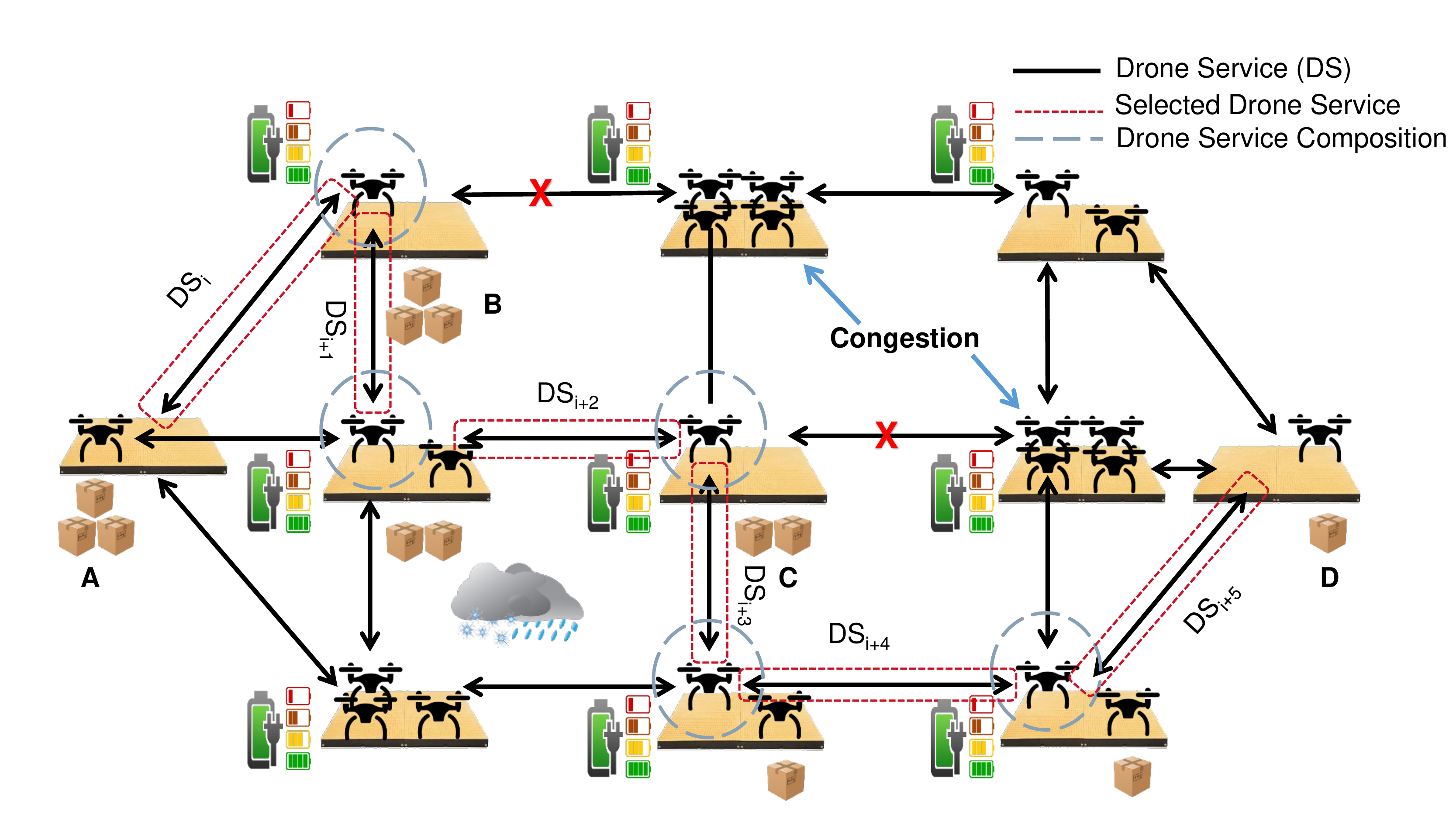}
    \caption{Drone service composition in a skyway network}
    
    \label{fig1}
\end{figure}

We leverage the service paradigm to reformulate the drone delivery in a skyway network as a \textit{drone service composition} problem. In the real world, customers have different QoS preferences for the service delivery \cite{10.1007/978-3-642-10383-4_6,5492672}. Therefore, we select and compose drone services with the best \textit{QoS}. Without loss of generality, we use the \textit{delivery time} as the main QoS. The \textit{composition} of drone services is the process of selecting and aggregating drone services leading to the destination \cite{9590457}. A drone service composition consists of (1). selecting the best linear composition of drone services to deliver the packages and (2). selecting the best strategy for recharging drones. Fig. \ref{fig1} represents a drone service composition scenario where three packages need to be delivered from point A to points B, C, and D. In some instances, a drone may not be able to fly from the source location to all destinations without recharging at intermediate stations due to battery limitations. The drone services, e.g., $DS_1, DS_2,\ldots, DS_m$, have their own QoS values for travel time and cost. Hence, different composite QoS values for time and cost may provide different service compositions. For example, two services, i.e., $DS_i$ and $DS_{i+1}$, may be composed to support the package's fastest delivery and avoid congestion at recharging stations. Each drone from a service provider is assigned a different delivery plan. Therefore, no \textit{handover} of packages is assumed to occur among drones flying in the same skyway network.

We propose a novel \textit{Service-Oriented Architecture} (SOA) for multi-package delivery using a single drone. The proposed architecture presents a high-level abstraction of the drone delivery system that is deployed using a skyway network. In this regard, several approaches exist that focus on optimizing the drone delivery system by proposing payload-mass-aware trajectory planning \cite{tsoutsouras2020payloadmassaware}, load-dependent flight speed-aware \cite{9052143}, and deadline-constrained \cite{BNCSS102} drone delivery. However, none of the aforementioned approaches considers the recharging conditions at intermediate stations and provides a high-level abstraction of the drone delivery system in the context of the service paradigm to offer optimal delivery solutions. We take into account the recharging constraints of drone-based provisioning of delivery to provide optimal solutions in terms of delivery and computation times.

We propose a heuristic based on the geographical area for service delivery and the recharging pad's availability at neighboring stations. In the first step, we constrain the search space using a graph-based heuristic to reduce the candidate drone services.
We use a Line-Of-Sight (LOS) directed from the source to all destination nodes for bounding the \textit{potential search space}. We then use a geometric method to determine the largest angle covering all the delivery targets. The LOS and the geometric method provide the radius of a subgraph to reduce the search space by eliminating \textit{non-optimal} nodes.
In the second step, we compute a time-optimal service composition plan for delivering packages to each delivery target, considering the services in the subgraph.
We conduct experiments to analyze our proposed graph-based heuristic approach's performance in comparison to the exhaustive drone service composition approach.
We define our main contributions as follows:

\begin{itemize}
    \item[$\bullet$] A novel service-oriented architecture for drone-based multi-package delivery.
    \item[$\bullet$] LOS and graph-based heuristics to constrain the search space for drone delivery service selection.
    \item[$\bullet$] A new drone service composition algorithm for multi-package delivery.
    \item[$\bullet$] Performance evaluation to show our proposed approach's efficiency.
\end{itemize}


\section{Motivating Scenario}
Drones deliver multiple packages within Texas, USA. Let us assume a drone service provider receives requests to deliver packages from \textit{San Marcos} to \textit{McQueeney}, \textit{Bulverde}, and \textit{San Antonio} with distances 46 km, 74 km, and 92 km from \textit{San Marcos}, respectively. A delivery drone can serve a maximum distance of 3 to 33 km depending upon the payload attached to it and the battery used.
A drone's flight range is determined using the payload weight, drone speed and battery capacity and its consumption rate. A drone may require a recharge multiple times to satisfy the delivery requests \cite{9284115,jermaine2021demo}. In this respect, it is crucial to avoid congested stations for time-optimal drone delivery.

We create a skyway network for flying drones following the Federal Aviation Administration regulations, including operating in LOS and avoiding no-flight and restricted areas. \textit{Each skyway segment is abstracted as a drone service}. The building rooftops within the Texas area represent the skyway network nodes.
Each rooftop is fitted with a fixed number of recharging pads for the drones to land and recharge. A congested station is one where all recharging pads are occupied. In this context, a station may become congested due to the arrival of delivery drones at the same time at a particular station \cite{10.1145/3460418.3479289}.
Avoiding the congested stations to plan deliveries would result in faster delivery services. We reformulate the delivery of packages using drones as the composition of a set of drone services at intermediate recharging stations.

\section{Related Work} \label{relatedwork}

Existing approaches to drone deliveries can be divided into two broad categories: (1) Data-oriented Approaches and (2) Service-oriented Approaches.

\textbf{Data-oriented Approaches:} Most of the existing data-oriented approaches formulate the drone delivery problem as TSP or VRP. In that regard, data-oriented approaches address the point-to-point package delivery problem using drones \cite{pinto2022point}.
The routing problem for delivery drones is presented as energy minimizing VRP \cite{10.1007/978-981-15-3514-7_91}. This study aims to find a route that consumes minimum energy to visit all customers. It is assumed that the drone's energy consumption depends on the travel distance and the payload weight. In addition, the drone's flight speed is assumed to remain constant irrespective of the changes in payload weight. A dynamic programming-based algorithm is presented to address the drone's routing problem. \textit{The proposed routing problem does not consider the LOS drone flying regulations to make deliveries, recharging constraints of drones, and congestion at stations.}

A deterministic automated drone delivery system is presented in \cite{choi2017optimization}. In the proposed system, the drones may carry multiple packages to deliver in a predefined geographical area.
A linear energy consumption function is used based on the payload weights and speed of the drones. This study aims at minimizing the total costs of operating an automated drone-based delivery system. The total costs include the operational costs, capital, and the costs incurred by the customer's waiting times. The study indicated that the drone service providers and customers are benefited from long hours of operation in areas with high demand densities. \textit{The proposed automated delivery system does not consider the LOS restrictions to fly drones and congestion at stations}.

A UAV scheduling problem is investigated to deliver packages from a warehouse to respective destinations \cite{8845427}. The objective is to maximize time-related customer satisfaction. A 0-1 knapsack algorithm is employed to find an initial solution. Then, a UAV Flight Mission Scheduling Framework is proposed that includes a local search algorithm and a simulated annealing algorithm to optimize the target. Simulation experiments validate the effectiveness of the algorithms for finding an optimal target. \textit{The proposed scheduling framework focuses on point-to-point service delivery in a given area ignoring the drone's recharging requirements.}

\textbf{Service-oriented Approaches:} Service-oriented approaches provide an \textit{elegant} and \textit{congruent} framework for the effective provisioning of service delivery using drones \cite{2021335,oosterhaven2018vehicle}. There has been an increase in studies that focus on the use of service-driven approaches to address the drone delivery problem, especially multi-package deliveries in heterogeneous environments.
A resilient drone service composition framework is presented for providing delivery services taking into account the congestion conditions at stations \cite{2021335}. The proposed resilient composition framework consists of a drone service model for the constraint-aware delivery services. In offline settings, a deterministic lookahead algorithm is used to generate an initial drone service composition. An adaptive lookahead service composition algorithm is developed to adopt the changes that occur in the offline service composition and compute a new local composition plan to meet the customer's delivery requirements. \textit{This resilient framework is limited to generate a service composition plan for the single package delivery using drones in one trip}.

A VRP with a single drone is studied to deliver more than one package in one trip in \cite{oosterhaven2018vehicle}. The geographic distribution of multiple warehouses is modeled to ensure the delivery of two packages in one trip. These warehouses serve as pick-up locations for packages and battery swapping stations for drones.
The nearest neighbor algorithm is proposed to find a delivery path for minimizing the total cost.
Battery swapping requires the precise landing of drones. Furthermore, a large number of batteries must be available at each station, specific to different types of drones. As the technology is yet to mature, it is unrealistic to find spare swapping batteries where thousands of drones are moving. In addition, the proposed nearest neighbor approach does not incorporate the congestion of drones at stations. To the best of our knowledge, this study is the first attempt to present an SOA for multi-package delivery using a single drone that takes into account congestion at recharging stations.


\begin{figure}
    \centering
    \includegraphics[width=0.4\textwidth]{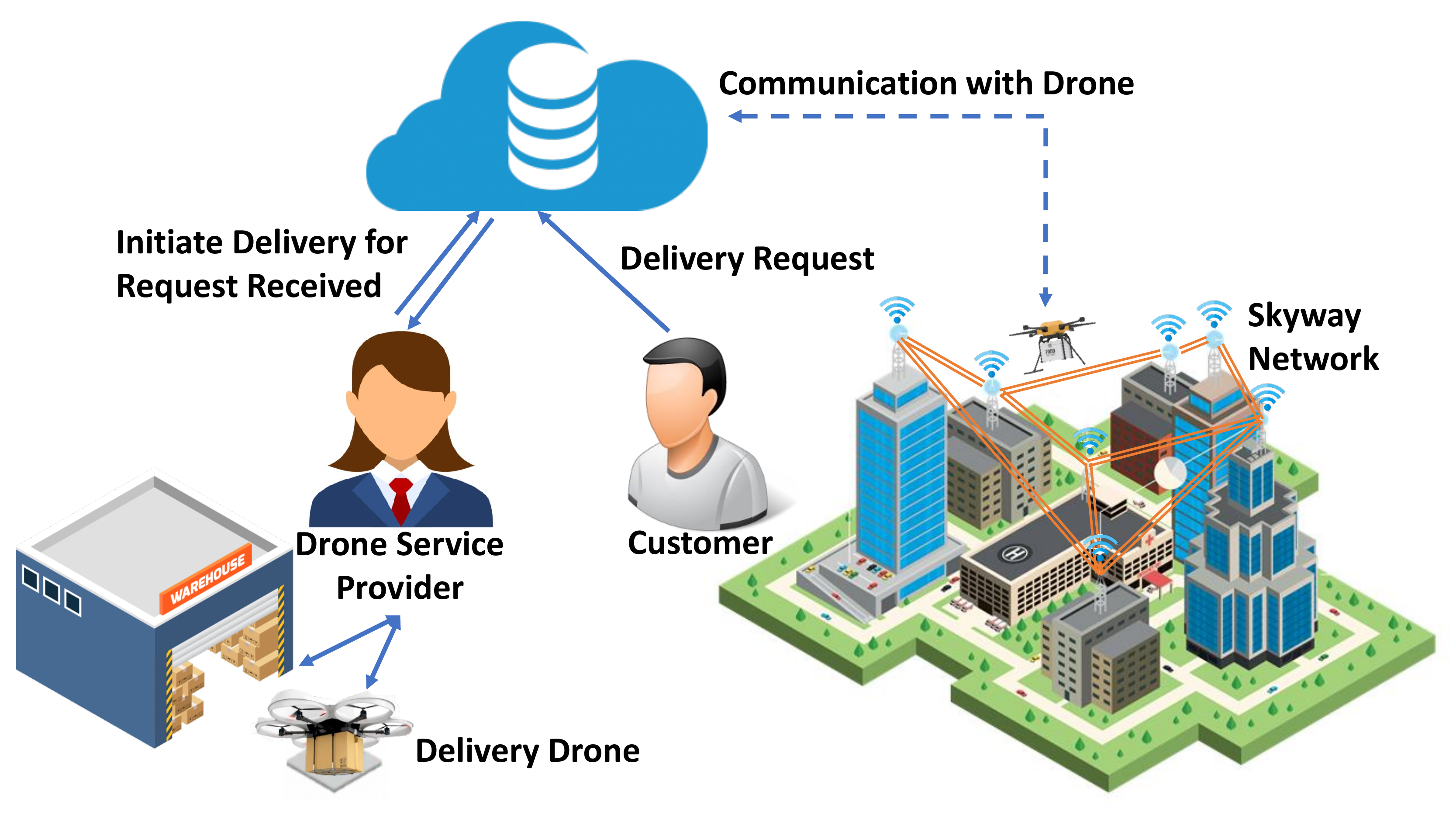}
    \caption{High-Level SOA Design}
    \label{fig13}
\end{figure}

\section{Service-Oriented Architecture for Drone-based Multi-Package Delivery}

We present an SOA for multi-package delivery using drones. The proposed SOA includes three main components: (1) High-Level SOA Design, (2) Skyway Network and Notations, and (3) Drone Service Composition Model.

\subsection{High-Level SOA Design}

This section describes a high-level design of the service-oriented drone delivery architecture. The proposed architecture is based on the skyway network deployment of drone deliveries, as shown in Fig. \ref{fig13}. The proposed SOA is an adapted architecture and includes a cloud-based drone delivery system where drone service providers are registered to offer delivery services. A customer selects a service provider and makes a delivery request through the system specifying the QoS attributes for the delivery. A \textit{message-based interaction protocol} facilitates the communication between the service provider and the customer \cite{alshinina2017performance}. The drone service provider initiates the delivery operation upon receiving the delivery request. In this regard, a drone is first selected from a set of available drones considering the QoS attributes specified by the customer. Then, the requested package is attached to the delivery drone and a drone service composition plan is computed from the source to the destination. The service composition plan selects the optimal drone services that meet the customer's requirements of minimum delivery time. Drones have limited processing power and battery capacity \cite{9590339}. Therefore, the heavy load computations cannot be executed on the drones.

We leverage a \textit{sensor-cloud infrastructure} to offload the processing load of drones to the cloud. The overall computations are carried out at two levels: \textit{drone level} and \textit{cloud level}. We perform \textit{lightweight} computations, including collection and analysis of sensor and battery data at drone level. However, \textit{heavyweight computations} including drone service compositions to constitute a skyway path from the designated source (i.e., warehouse) to the delivery destination (i.e., customer's building) are computed at the cloud level. The drone communicates with the cloud to store and retrieve real-time data during delivery. In addition, the cloud sends a set of instructions to the drone, including picking up the package, initiating the delivery operation, waiting and stopping at recharging stations, and delivering the package to the destination. The customer receives a notification from the drone service provider when the package is successfully delivered.

\section{Heuristic-based Drone Service Composition} \label{problemdefinition}


We use a graph-based heuristic to create a subgraph. In this respect, a subgraph is defined as a subset of nodes and edges of the spatio-temporal graph $G$ (i.e., skyway network). This subgraph covers all the destinations of packages assigned to the selected drone. We eliminate the rest of the nodes and edges of the graph $G$ to reduce the search space. For example, a skyway network covers 50 km of a given geographical area. The farthest destination of a drone is within 10 km from the source location. Our graph-based heuristic reduces the search space by eliminating the services that have a distance of more than 10 km from the source location. In contrast, the exhaustive composition takes into account all the drone services to compute the optimal service composition. This approach is computationally expensive and time-consuming for large-scale instances. The graph-based heuristic provides computationally efficient solutions compared to the exhaustive composition approach.

\begin{algorithm}[t]
\small
\caption{Graph-based Drone Service Composition}\label{alg:algorithm1}
	\begin{algorithmic}[1]
    \REQUIRE
		$G$, $d$, $src$, $dsts$
	\ENSURE
		$CompDS$\\
		\STATE $CompDS \gets \phi$
		\STATE compute LOS from the source to all destinations
        \STATE $subgraph\_nodes \gets$ compute subgraph using LOS to each destination
		\FOR{each $dst \in dsts$}
		\STATE $path \gets $ compute\_path ($subgraph\_nodes$, $src$, $dst$)
		\ENDFOR
		\STATE $cost\_matrix \gets$ drone $d$ delivery time for each path based on distance
		\STATE $CompDS \gets$ find best composition plan for drone $d$ in subgraph based on the cost matrix
		\RETURN $CompDS$
	\end{algorithmic}
\end{algorithm}

\subsection{Graph-based Heuristic Algorithm}

We present a new graph-based heuristic algorithm to restrict the search space and optimize the selection and composition of drone services. We extract potential nodes and edges from the skyway network, including all the delivery targets for a selected drone. We compose skyway segment services to deliver packages in minimum delivery time. In this regard, we select and compose only one drone service at a time at intermediate recharging stations. Algorithm \ref{alg:algorithm1} describes our proposed graph-based heuristic approach. The output is a right drone service composition plan from a given source to all destinations for the input of the network as a spatio-temporal graph $G$, the selected drone $d$, the source $src$, and the destinations $dsts$. We create an empty list for drone service composition plan $CompDS$ (Line 1). We consider a virtual LOS route to each destination node (Line 2). We then find a subgraph using LOS to each destination that includes all nodes between the two farthest destinations (Line 3). The shortest path service composition to each destination is then computed (Lines 4-6). We generate a cost matrix of delivery times required to traverse each skyway segment for the selected drone $d$ (Line 7). The delivery time includes travel time from one node to the next and waiting and recharging times at intermediate recharging stations. We finally compute and return the best drone service composition plan based on the cost matrix (Lines 8-9).

\section{Performance Evaluation}
We use the following settings to analyze the performance of our proposed graph-based approach for the selection and composition of drone services:
\begin{itemize}
    \item[$\bullet$] \textbf{Performance Metrics:} The delivery time is the most important factor in efficient service delivery. We use (1) \textit{delivery time} and (2) \textit{execution time} as performance metrics. The runtime complexity of the algorithms is analyzed based on the execution time.
    \item[$\bullet$] \textbf{Baseline:} We compare our proposed graph-based service composition approach with exhaustive service composition approach.
\end{itemize}




\subsection{Experiment Settings}

We use the NetworkX Python library to construct a skyway network for drone deliveries. Each node is modeled as a station with fixed number of recharging pads or a target to deliver packages. Further, we model multiple drone delivery services to operate in a skyway network along with their quality parameters, e.g., payload and battery capacities. We conduct a set of experiments where each experiment is run for 50\% times the total nodes in the skyway network. For example, we perform an experiment 15 times if the network contains 30 nodes. A random source node and random destination nodes are selected for each experiment.
All the algorithms are written in Python.
The details of experiment parameters are given in Table \ref{tab:table5}.





\begin{table}
\centering
\caption{Experiment Parameters}
    
\label{tab:table5}
\begin{tabular}{|l|l|}
\hline
 \textbf{Parameter} &  \textbf{Range of values} \\

\hline

Multi-package delivery drone &  DJI Matric 300\\ \hline

Max payload & 15.3 Kg \\ \hline
Max flight time & 55 min \\ \hline


Max speed & 82.8 km/h \\ \hline

Recharging time & 2.15 hours \\ \hline

Max skyway network nodes &  35  \\ \hline


Recharging pads at each station & 4 \\ \hline

Experiment run & 50\% \\

\hline
\end{tabular}
\end{table}

\subsection{Results and Discussion}

Our graph-based heuristic approach constrains the search space and composes the right set of drone services to deliver the packages faster.

\begin{figure}[t]

    \centering
    \includegraphics[width=0.4\textwidth]{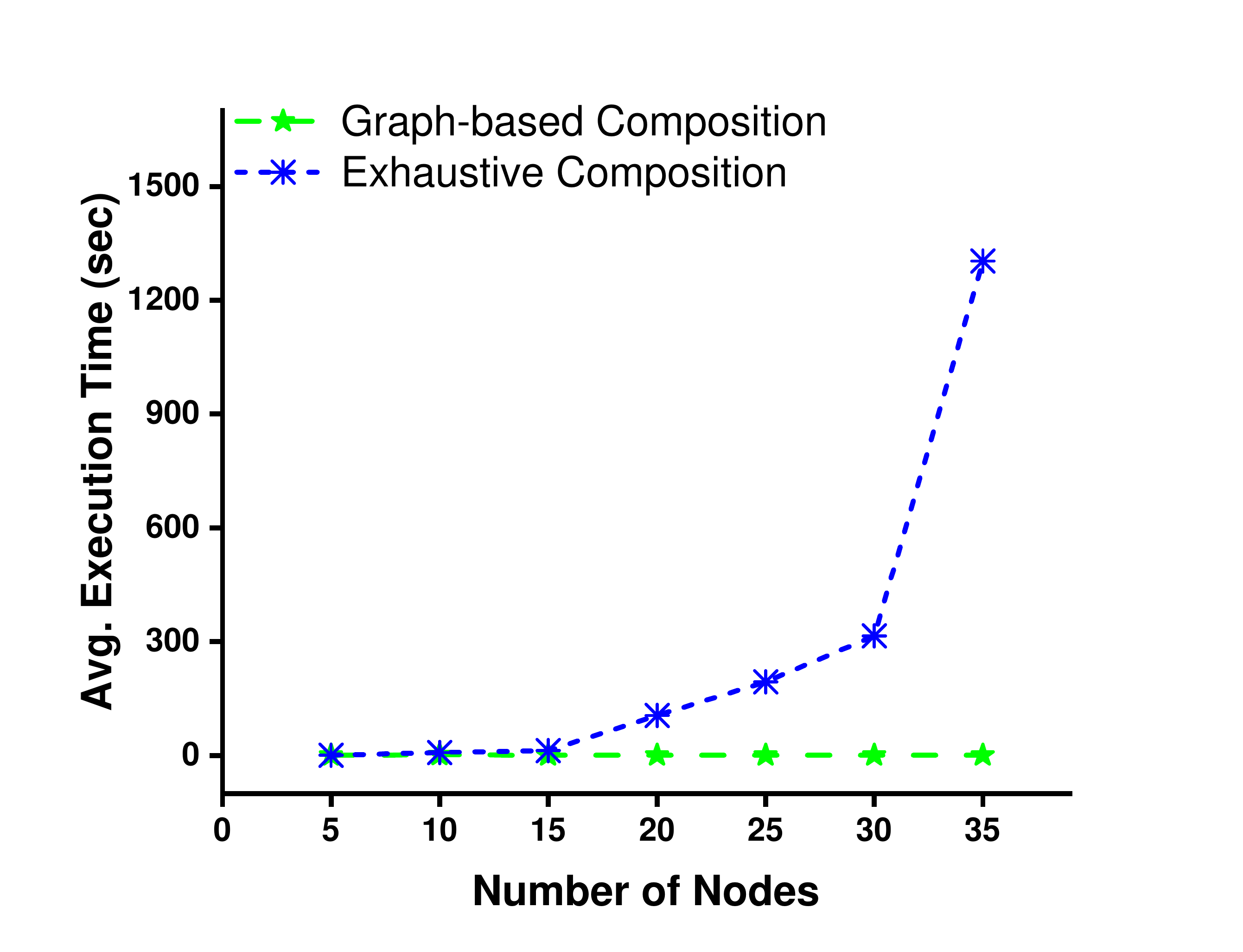}

    \caption{Average Execution Time for Varying Number of Nodes}

    \label{fig4}
\end{figure}





\subsubsection{Average Execution Time}
The computational complexity is critical to assess the efficacy of an algorithm. Therefore, we evaluate the algorithm's computational complexity in terms of execution time. The exhaustive composition approach is computationally expensive as it takes into account all the skyway network nodes for service compositions. The execution time is directly proportional to the number of possible compositions. We evaluate the execution times for the varying number of nodes. The average execution times for our proposed graph-based heuristic approach and exhaustive composition approach are illustrated in Fig.~\ref{fig4}. The execution time of graph-based approach is significantly less than the baseline approach. The exhaustive composition execution time grows exponentially when the network size increases. The results show that the baseline composition approach is not applicable in real-world settings due to its exhaustive behavior.





\begin{figure}[t]

    \centering
    \includegraphics[width=0.4\textwidth]{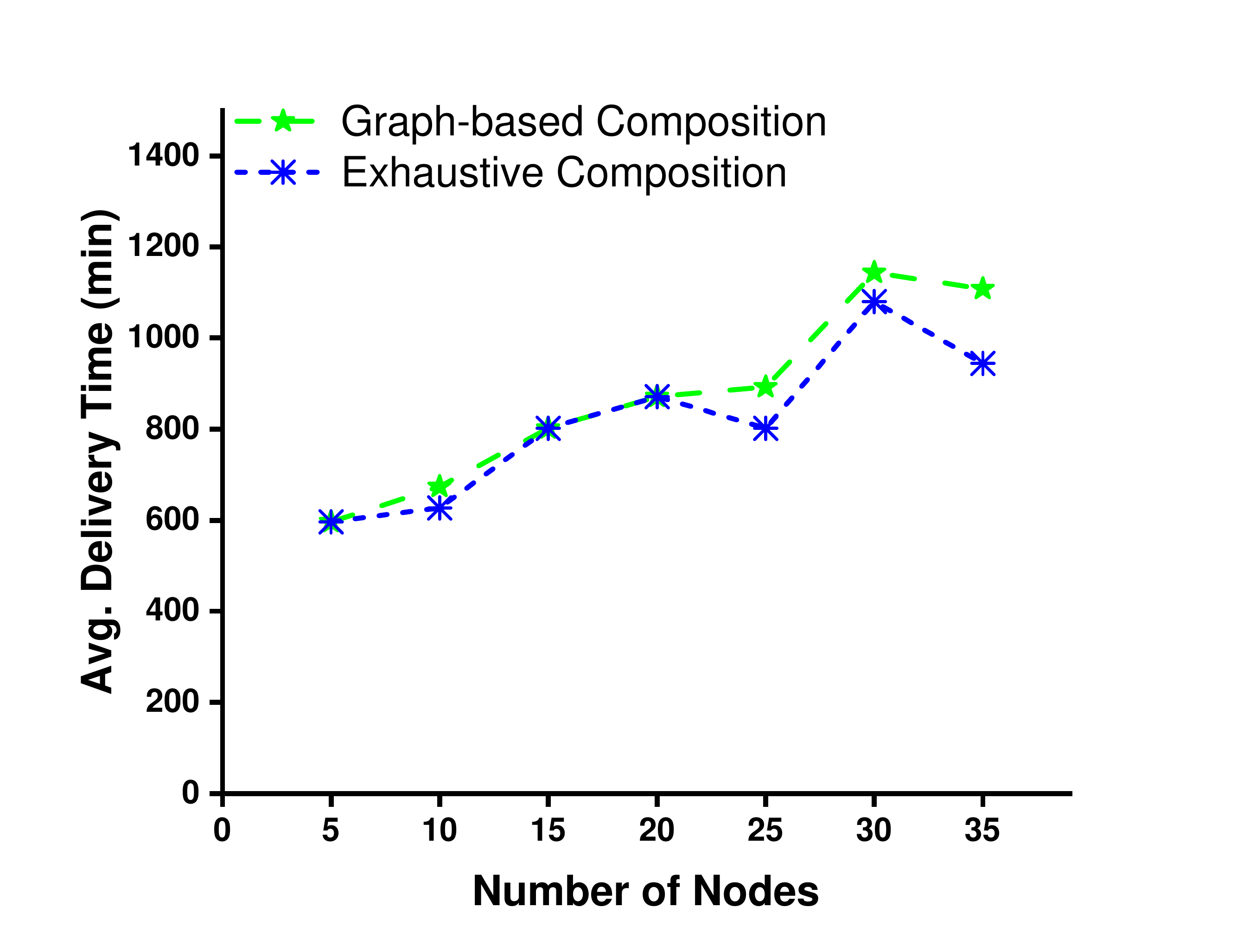}

    \caption{Average Delivery Time for Varying Number of Nodes}

    \label{fig7}
\end{figure}









\subsubsection{Average Delivery Time}
A typical delivery drone can carry multiple packages at the same time \cite{s22052045}. However, the total number of packages a single drone can lift depends upon the weights of the packages and the drone's payload capacity. We evaluate the delivery time of packages using a single drone for the varying network sizes, i.e., the total nodes in a network. The delivery time to visit all destinations includes travel time from one node to another, waiting time, and recharging time at intermediate nodes (i.e., recharging stations). The drone's payload weight, speed, and congestion at stations mainly affect the overall delivery time. Fig. \ref{fig7} presents the delivery times for our proposed graph-based heuristic and exhaustive composition approaches. Our proposed graph-based heuristic approach provides near-optimal solutions in comparison to the exhaustive composition approach.

\section{Conclusion}

We propose a novel service-oriented architecture for multi-package delivery using a single drone. We model the service as a skyway segment in a network that is served by a drone. We propose a graph-based heuristic approach for the selection and composition of a right set of drone services to deliver multiple packages using a single drone in one trip. We perform several experiments to compare our graph-based heuristic approach with the baseline composition approach. We observe that our proposed graph-based heuristic approach is computationally efficient in comparison to the exhaustive composition approach. In the future, we plan to investigate the impact of a varying number of destinations and extending the subgraph to include more nodes on the delivery time.

\section*{Acknowledgment}
This research was partly made possible by LIEF project (LE220100078) grant from the Australian Research Council. The statements made herein are solely the responsibility of the authors.

\bibliographystyle{IEEEtran}
\bibliography{references}

\end{document}